# The CMS ECAL Trigger and DAQ system: electronics auto-recovery and monitoring

Prasanna K. Siddireddy on behalf of the CMS Collaboration

*Abstract*—In 2017, the Large Hadron Collider (LHC) at CERN has obtained an astonishing result providing proton-proton collisions with an integrated luminosity of about 50 fb-1. The Compact Muon Solenoid (CMS), a general-purpose detector operating at the LHC, has been able to record 45.39 fb-1. In this frame, ECAL, the CMS Electromagnetic Calorimeter based on PbWO4 crystals, has shown a good performance with a very stable DAQ system even during higher LHC luminosity peaks. The ECAL DAQ system follows a modular schema: 75848 crystals are divided in sectors (FED), each of them controlled by 3 interconnected boards. These handle the Front-End electronics configuration and control, compute the Triggers Primitives for the CMS Level1 Trigger and read out the data. During the data taking, Front End electronics is prone to occasional errors, induced by particles interactions. If these errors are not handled appropriately they can cause data loss in the section affected or even block the data taking of the experiment. To prevent these situations, an automatic recovery mechanism has been developed in the ECAL DAQ software. The same software dedicated to the configuration of the boards has been modified to check periodically their status. Depending on the level of the error, the software can trigger a reconfiguration of a single component, of the full board or even mask the affected section and exclude it from the following run. Under these circumstances it is crucial to provide to the experts a real time view of the electronics status. Possible errors and corresponding automatic actions have to be shown, as well as logged, in order to have a clear picture of the status of the detector and of its general efficiency. For this purpose, a web application running on a light JavaScript server framework based on Node.js and Express.js, has been developed. It is composed by several routines that cyclically collect the status of the electronics and expose the information to web requests. On client side, graphical interfaces, based on Vue.js libraries, display the information of the electronics status and errors. Server-side routines store electronics errors information in a SQLite database in order to perform offline analysis about the long term status of the boards.

## I. INTRODUCTION

INSIDE the CMS (Compact Muon Solenoid), the proton beams accelerated by the CERN LHC (Large Hadron Collider) collide every 25 ns creating about 40 inelastic interactions at every BX (bunch-crossing). CMS [1] is a multipurpose detector designed for the precision measurement of leptons, photons, and jets, among other physics objects, in proton-proton as well as heavy ion collisions at the LHC [2]. The enormous amount of data generated at 40 MHz is beyond the storage capabilities: the DAQ (Data Acquisition) and triggering system is in charge of the real-time data collection and selection at every BX.

The short time interval of 25 ns between each BX represents a strict requirement for the detector response and electronic readout: the effect of the pile-up can be reduced by using high granularity detectors with good time resolution and large number of channels that need a very good synchronization. The cross-sections of processes studied by LHC, such as Higgs production, are more than 9 orders of magnitude lower than the elastic pp (proton-proton) cross-section, so the trigger system represents the start of the event selection process in order to discard only non-interesting data.

The CMS trigger utilizes two levels [3]. The L1 (Level-1) Trigger is composed by custom designed programmable electronics which collects information from the subsystems of CMS and implement selection algorithms to reduce the acquisition rate to 100 kHz at within 3.8 µs [4]. Then the chosen events are further filtered by the HLT (High-Level Trigger) [5], a software application able to partially reconstruct the event in order to apply more refined selections. This reduces the rate to about 1500 Hz [8].

## II. ELECTROMAGNETIC CALORIMETER

The ECAL (Electromagnetic Calorimeter) is the CMS sub-detector designed to measure the energy of electrons and photons produced in the collisions. ECAL comprises of EB (Electromagnetic Barrel), EE (Electromagnetic Endcap), and ES (Endcap Preshower). However, in the discussion that follows ECAL is used to refer to just EB+EE. ECAL is a hermetic, homogeneous calorimeter comprising of 75848 PbWO$_4$ crystals with 61200 of the crystals mounted in the central barrel part, closed by 7324 crystals in each of the 2 endcaps. The calorimeter is in the form of a cylinder as can be seen in figure 1 and it uses η-ϕ coordinate system, η being the pseudorapidity and ϕ being the azimuthal angle [3].

 The DAQ system of ECAL is made by an on-detector part and an off-detector one. The on-detector FE (Front End) electronics reads the analog signal from the photodetectors, it amplifies and digitizes it, and then it calculates the information required by the L1 Trigger, called TP (Trigger Primitives), that is the total energy deposited in every TT (Trigger Tower). This information is sent to the off-detectors boards, while

The author Prasanna K. Siddireddy is a graduate student at University of Notre Dame. He is currently based at CERN working on the CMS experiment. (email – psiddire@nd.edu)

measurements relative to each crystal are held in buffers in the FE waiting to be selected and read out.

The ECAL off-detector electronics [6] is made by 54 identical modules called FEDs (Front End Drivers). Every FED is made by 3 interconnected boards: The TCC (Trigger Concentrator Card), the CCS (Clock and Control System), the DCC (Data Concentrator Card). The TP calculated in the FE are elaborated by the TCCs and sent to the L1 Central Trigger. If the event is considered interesting, an L1 Accept signal is sent by the central system to all the detector parts: in ECAL the L1A is redistributed by the CCS board. At this point the DAQ process starts: the DCC board reads the crystals data from the FE with full granularity and it reduces the payload of the event data to be saved, thanks to additional information about the significance of each TT processed by a secondary circuit, the SRP (Selective Readout Processor) [7].

The ECAL DAQ system is handled by a distributed software that is in charge of the configuration of all the electronic boards and of the life cycle of the acquisition process. The tree control structure makes the whole system particularly modular: the operator, through a web interface called the FM (Function Manager), controls the ECAL Supervisor, an application that handles a Supervisor for every off-detector board which is responsible for the configuration of the electronics.

### A. L1 Calorimeter Trigger

The L1 Calorimeter Trigger receives information from the ECAL, HCAL (Hadron Calorimeter), and HF (Forward) calorimeters of the CMS detector, and uses it to reconstruct electron, photon, tau, and jet candidates, and to compute energy sums. The input is organized in TT, each corresponding to a region of approximately $0.087 \times 0.087$ of extension in $\eta$, $\varphi$ in the central part of the detector. A slightly more complex geometry is used in the circular endcaps of the detector. Each TT encodes the energies deposited in ECAL, HCAL and HF at a specific position in the detector [11].

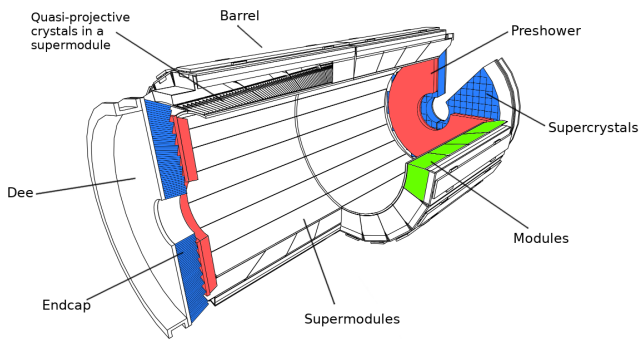

Fig. 1. The EB has an inner radius of 129 cm. It is structured as 36 identical "Supermodules," each covering half the barrel length and corresponding to a pseudorapidity interval of $0 < |\eta| < 1.479$. The EE at a distance of 314 cm from the vertex and covering a pseudorapidity range of $1.479 < |\eta| < 3.0$, are each structured as 2 "Dees" [3]

The L1 Calorimeter Trigger is designed to access the entire detector information at the TT granularity. A TMT (Time Multiplexed Trigger) [9] architecture, structured in two separate processing layers, is used. The data are collected from the calorimeters in the Layer-1, which distributes data of adjacent bunch crossings to one out of nine processing nodes of the Layer-2. In this way, each Layer-2 card receives the information from all the TT in the event. An additional redundant node is also available to be used in case of failure of any other node. The outputs of these nodes are collected by a de-multiplexing (demux) node, are ranked by transverse momentum, and are sent to the µGT (micro Global Trigger) [8].

TP from ECAL and HCAL are combined at Layer-1. The TP are used to reconstruct electrons candidates with a dynamic clustering algorithm. Combination of EG-like (Electron Gamma) clusters (including HCAL TP) are used to reconstruct tau lepton candidates and a 9x9 TT sliding window approach is used to build jet candidates. Multiple working points for isolation, shape vetoes, the fine grain bit and the ratio of the H/E is used to identify electromagnetic energy deposit. The final trigger decision at Level-1 is performed at the µGT level which then distributes the L1 Accept signal [10].

The L1 Accept decision is communicated to the different sub-detectors through the TTC (Timing, Trigger and Control) system. Every BX has to be analyzed but the L1 Trigger requires some time to take its decision: the total time allocated for the L1 decision and the transit of the signals from the on-detector electronics to the services cavern is about 3.8 µs. During this time, the detector data must be held in buffers while trigger data is collected from the FE electronics and processed off-detector.

When the L1A (L1-Accept) signal is received by the system, the detector high-resolution data of the selected event is transferred to the FED cards in the counting room with optical links, operated at 800 MB/s. Every subsystem of CMS has then different electronic boards for signal processing, zero-suppression, data-compression, and calibration. The processed data is contained in several hundred FE readout buffers, ready to be elaborated by the HLT.

The cDAQ (Central DAQ) system is in charge of collecting the elaborated data from 650 sources at the nominal L1 trigger rate, combining every event fragment from the different sub-detectors into a unique packet that is then sent to the HLT farm. Each data source provides event fragments of 2 KB/s; thus, the average throughput of the system is 500 GB/s (depending on the L1 trigger rate).

### B. The Trigger Path

The basic detector unit is the TT, a group of 25 adjacent crystals. Every TT is controlled by a group of boards called the FE that process at 40 MHz the crystal data received from the sensors. The FE electronics is the first step of both the Trigger and the DAQ path. The FE consists in a motherboard, 5 VFE (Very Front End) boards and one FE board. Each VFE reads the data from 5 crystals each with a MGPA (Multi-Gain Preamplifier) [12] (with gains 1,6,12), an ADC 40 MHz digitizer and radiation-hard buffers to send the data to the FE. An integrated logic selects the highest non-saturated signal as output and reports the 12-bit of the corresponding ADC with two bits encoding the ADC channel number (to store the gain used). If the read-out switches to a lower gain as the pulse grows, the return to higher gain is delayed by five samples after the threshold is re-crossed.

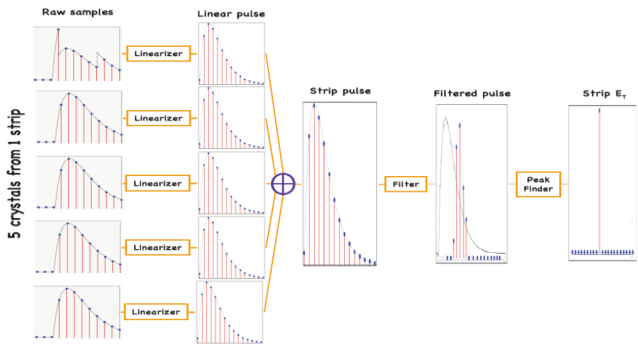

Fig. 2. The ECAL trigger primitive algorithm [13]

The ECAL TPG logic is implemented in the FE cards. This logic must reconstruct the signal amplitude to be assigned to each BX from the continuous stream of successive digitization's. Each VFE card receives the input from 5 crystals amplified and digitized by the MGPA and ADC. The 5 digital signals from the crystals of a strip (5 crystals of TT in $\phi$) are first linearized. The linearization process takes into account the gain used in the readout electronics and corrects the ADC counts by the calibration coefficients. A multiplicative factor is applied to get linearized ADC counts proportional to the transverse energy deposit in the crystal. The 5 linearized signals of the crystal are then summed up to produce the strip signal. An amplitude filter is applied to measure the amplitude of the strip pulse. The filter is based on linear weighted sums where the weights take into account the expected shape of the signal and subtract dynamically a possible residual pedestal.

The output of the amplitude filter is then filtered by a peak finder stage that keeps only the maximum as a measure of the $E_T$ contained in the strip. The signals of the 5 strips of a tower are finally summed up to provide a measurement of the total $E_T$ of the TT. The dynamic range of the $E_T$ must then be reduced from 10 bits to 8 bits. Look-up tables are used for this purpose. The look-up tables are chosen so that the loss of resolution due to the compression follows approximately the intrinsic ECAL resolution. In parallel, the signals of the strips are combined in a Fine Grain filter producing 1 bit, the fine grain veto bit, indicating the transverse extent of the electromagnetic energy deposit. In summary, the ECAL TPs are made of the total $E_T$ of the TT encoded on 8 bits, accompanied by the fine grain bit. They are computed for all TTs and for each BX.

The trigger data elaborated by the FE are sent at the LHC clock to the off-detector electronics. There the TCC [14] collects data from 68 TTs in the barrel, corresponding to a super module and, in the End-Caps, up to 48 pseudo-strips (5 pseudo-strips per End-Cap FE boards). In the endcap, 2 TCC boards cover a 20° sector. Hence, there are 4 TCC boards per 40° sector which correspond to one ECAL FED. The TCC completes the TP generation and sends the data to the L1 through the SLB (Synchronization and Link Board) at each BX.

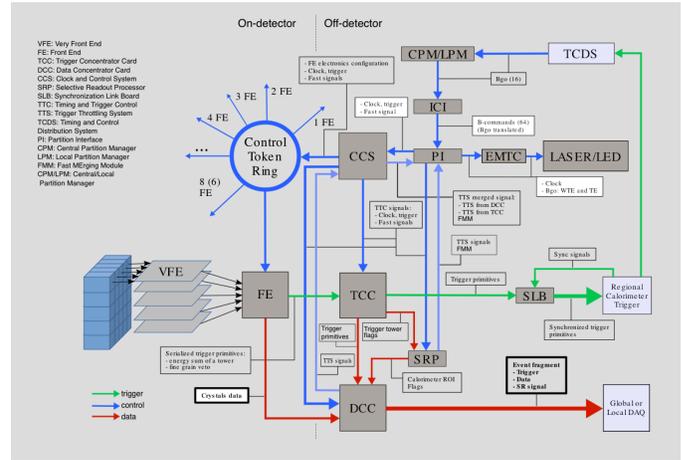

Fig. 3. A full schematic of the ECAL DAQ's on-detector and off-detector electronics showing how the FE, CCS, TCC, and DCC intercommunicate with each other and with RCT and Global DAQ

The different particles in the detector introduce non-negligible differences on the TP arrival time in the different processors. The SLB is in charge of the correct synchronization procedure that relies on the Bunch Crossing Zero (BX0) broadcast command from the central TCDS (Trigger Control and Distribution System). Finally, the TCC stores the TP during the L1 latency for subsequent reading by the DCC board.

*C. The Control Path*

The synchronization of the on-detector and off-detector boards with the LHC bunches is essential to assign the data to the right BX. The on-detector electronics must be carefully synchronized with the off-detector one. The CMS Global Trigger distributes the L1A signal through the TCDS system. This system includes the TTC and the TTS (Trigger Throttling System) and it is organized in several partitions that reach every sub detector of CMS.

The TTC system distributes the central clock of the CMS, aligned with the LHC clock, along with L1A signals and other special messages to take special triggers for calibration and tests. The TTS system is used in order to limit the trigger rate in case buffers of the electronics may overflow. The TTS receives inputs from the off-detector electronics about the status of the electronics and sends them to the global trigger to regulate the L1A signals emission.

The interlocutor of the TCDS system on the off-detector side is the CCS board. It distributes the clock and TTC commands to several FE boards using token rings [15], dedicated optical links. Moreover, it is connected to the TCC and DCC board from the backplane of the crate to which it sends the clock and control signals. Finally, the CCS board merges trigger throttling signals coming from TCC and DCC and forwards this information to the central TTS system.

*D. The DAQ Path*

The DAQ path starts when the L1A signal is broadcasted by the CCS to all the boards. The DCC is the main ECAL read-out unit and it is responsible for collecting crystal data from up to 68 FE boards in the barrel case [16]. The specific task of the DCC is the readout of the full-granularity crystal data upon a L1A decision. Every FE card receives the L1A signal and sends the raw crystal data corresponding to the right BX to the DCCs through optical links.

Also, the TCC boards receive the L1A and send their TP data for each TT to the DCC boards through the backplane. The DCC merges the crystal data with TP bits to save all the information elaborated in the Trigger Path. Moreover, the DCC performs and extensive integrity check of the data received from the FE: synchronizations errors and formatting errors are propagated online through the TTS. A buffer protection system is also implemented through TTS signals to inform the central trigger about busy state of the DCC board that inhibits new triggers or reduces trigger rate.

About 100 kB per event have been allocated for ECAL data, but the full event payload, if all channels are read out, exceeds this target by a factor of nearly 20. Reduction of the data volume, selective readout, can be performed by the SRP so that the suppression applied to a channel takes account of energy deposits in the vicinity, with the goal of achieving the best resolution for large energy deposits.

The calorimeter regions are classified by the TCC in three groups (low, medium and high interest) comparing the transverse energy deposited in each trigger tower with two configurable thresholds. Upon L1A signal these flags are transmitted to the SRP system which determines the readout mode that the DCCs have to apply to each trigger tower.

There are two readout modes: full readout and zero suppression. With full readout, the DCC saves the crystal time samples without any suppression. With zero suppression, the DCC calculates the amplitude of the signal for each crystal applying the FIR filter over the digitalized time samples and compares it with a configurable threshold: if the amplitude is above the suppression level the crystal data is saved, otherwise it is discarded.

After each the DCC has read the data, applied the selective readout and added the TP bits from the TCC, it sends the data packet to the central DAQ that collects all the event fragments to be delivered to the HLT software.

## III. ECAL TRIGGER AND DAQ - SOFTWARE

### A. The Configurable Database

The operation of the ECAL Trigger and DAQ system requires the configuration of parameters. The run conditions and the configuration used are necessary for the data-reconstruction process, moreover the full system configuration must be reproducible a-posteriori for debugging and verification purposes. So, a database is used as a long-term storage.

The configuration parameters have been sorted into three groups identified by a configuration key according to their function and update interval. [19]

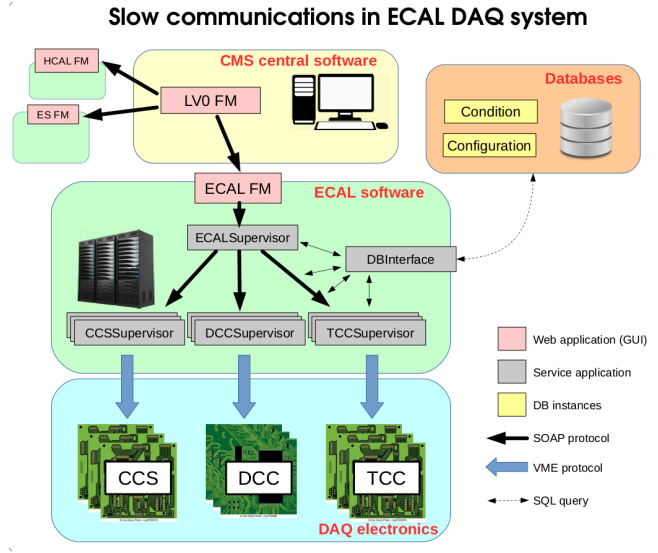

Fig. 4. At the top level, we have the L0 FM controlled by the operator who also controls the ECAL FM. The ECAL FM communicates with ECAL Supervisor which in turn communicates with the Resource Supervisors controlling the different DAQ electronics

The operation mode of the system is defined by the RUN_KEY associated with the ECAL_RUN_CONFIGURATION table. This defines the source of L1A triggers, the type of data reduction algorithm to be applied, the latency of the data and trigger path, the FE gain. The RUN_KEY specifies also the configuration of the different system resources and the type of the run: PHYSICS for collision runs, COSMIC for cosmic ray runs.

The FE_KEY defines all the parameters necessary to the DAQ path: the values of the MGPA DAC, the FE time alignment registers, the zero-suppression FIR weights and the list of channels to be excluded from the read-out.

The TPG_KEY defines the parameters for the TPG: the crystals linearization coefficients, the pedestal-subtraction coefficients, the TP FIR weights, the fine grain bits transverse energy thresholds and the list of crystals and towers to be excluded from the TP generation.

### B. The Online Software

The ECAL online software is responsible for the configuration, operation, and monitoring of the ECAL Trigger and DAQ system. The software has been developed as a modular distributed system based on web technologies and network communication protocols. The system accomplishes several tasks: it configures the hardware according to the parameters in the configuration database; it configures the ECAL calibration monitoring system; it monitors the hardware during data taking and reports error conditions to the operators; it integrates the ECAL sub-system in the CMS TriDAS system.

The base for all the CMS DAQ application is a C++ framework called XDAQ which provides the basic services needed by the online DAQ software as: an inter-process communication with SOAP-based channels, a data transport service based on the I2O binary protocol, a vendor independent database access service, a distributed logging, monitoring and error-reporting infrastructure, and a hardware access library.

The ECAL software has a hierarchical organization. At the top level, there are the FM, web application based on the RCMS (Run Control and Monitoring System) Java framework [18]. The FMs are the most external layer of the system, exposed to the action of the operator and responsible for managing the life-cycle of the lower layer applications.

The operators configure all the CMS TriDAS through the Level-0 FM which controls a FM for each subsystem. The configuration keys for each run are set in the FM as other resources like the data transfer to Tier-0 configuration. The different subsystems could be enabled or not in the Level-0 FM in order to do several system tests. Moreover, all the problems and error conditions that happened in the subsystems are delivered to the FM for a ready operator's response.

All the operations performed on the ECAL DAQ system originate in the ECAL FM and are propagated via SOAP-messages to the lower software layers. The building blocks of the system are the ECAL Supervisor and the Resource supervisors. The ECAL Supervisor is responsible for the coordination of the different DAQ resources for the electromagnetic calorimeter: it receives commands from the ECAL FM and dispatches them to Resource supervisors.

For each component of the hardware of ECAL, dedicated applications, generically called Resource Supervisors have been implemented. These receive the configuration parameters from the database according to the directives coming through the ECAL Supervisor and are responsible for the configuration and monitoring of the hardware resources. The set of operations performed on the hardware is fully specified by the list of detector components to be operated (FED_VECTOR) and the list of configuration keys chosen in the Level-0 FM. The software runs on a cluster of rack-mounted PCs located in the CMS service cavern. Eighteen machines are connected to the VME crate controllers and host the resources supervisors in charge of operating the Readout Units.

*C. The ECAL Supervisor*

The ECAL Supervisor application is responsible for the configuration of all resources controlled by the ECAL Online Software: all the VME readout-units, the SRP, the calibration monitoring system and the TTC interface. The ECAL Supervisor receives the configuration keys and the FED_VECTOR from the ECAL FM and propagates them to the other linked applications. During the Initialize state transition, a map of the system is built with the information collected by the different Resource Supervisors. Applications controlling the same type of resources are organized in groups called services and detector components are mapped to each application. A global list of resources is exported to the ECAL FM.

During the Configure state transition, the application determines which services need to be operated, based on the content of the RUN_KEY and also determines which applications to target, based on the requested FED_VECTOR. At the Start transition, the ECAL supervisor receives the run number from the ECAL FM, distributes it and creates a record in the CMS Condition Database to log the set of configuration keys and the FED_VECTOR used. This ensures that the system configuration used in any run can be reproduced at any time. Each type of VME board in the system has an associated Resource Supervisor.

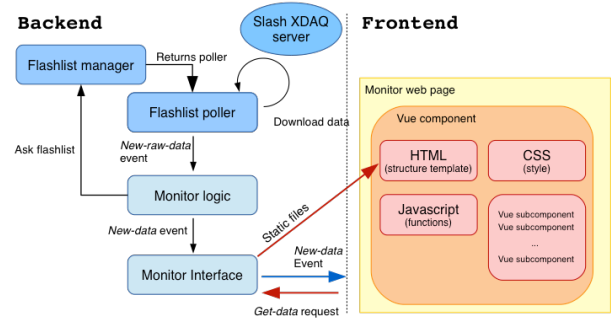

Fig. 5. A schematic showing the ecalView which is composed of Backend which is a Node.js server, an asynchronous event driven JavaScript runtime and a Frontend which uses Vue.js, a JavaScript framework for building user interfaces

The system comprises: 54 DCCSupervisor applications each controlling 1 DCC board, 18 CCSSupervisor applications each controlling 3 CCS boards, 54 TCCSupervisor applications each controlling 1 TTC board, 1 SRPSupervisor application controlling the SRP system, 1 LaserSupervisor application controlling the calibration monitoring system. All these applications implement the common FSM (Finite State Model) and provide monitoring data for the controlled resources.

IV. MONITORING

The ecalView is a web application made to monitor the status of ECAL online. Its purpose is getting monitoring data from XDAQ [17] slash server and elaborate it in order to offer useful information for ECAL shifters. It not only shows the data, but its final goal is creating tools on top of the data to help recognizing errors and reacting quickly to them. It is highly modular and easily extensible. It is possible to add new monitors in the ecalView framework in a simple and maintainable way.

Each monitor is composed by some standard files. For the backend, we used Node.js [20] which is an asynchronous event driven JavaScript runtime. For the frontend, we used Vue.js [21] which is a JavaScript framework for building user interfaces. Every monitor needs to follow this structure to be integrated in the ecalView infrastructure. The monitor gets the raw data from a component called flashlist- poller. Every few seconds (configurable) the poller downloads the data from the XDAQ slash server and sends it to the monitor-logic component via a "new-raw-data" event. Every poller can serve more than one monitor-logic (for example different monitors could use the same data). So, the pollers are loaded and configured centrally by the flashlists-manager component.

The monitor-logic registers itself to the event-emitter (the object inside the poller that sends the events) of the required poller through the flashlists-manager. The work of the monitor-logic module is to transform the raw-data coming from the flashlist-poller and make it available in the right format for the rest of the software. For example, it extracts the errors from all the channels (TT) of a board and creates a short list that can be displayed by the monitor as a summary. The monitor-logic starts its processing on a "new-raw- data" event from the poller and at the end sends out a "new- data" event to all the listeners. At this point, other components can use the new elaborated data from the monitor-logic: for example, the monitor-interface can transmit it to the web page.

The monitor-interface component exposes the monitor resources to the outside world. It contains the endpoints (REST API) for monitor data and metadata and it serves the static files needed to build the monitor web page. The monitor-interface has a special relationship with the JavaScript component that handle the front end: it sends out "new-data" events and other messages to keep the interface up-to-date. Under the hood the monitor-interface is an Express Router and serves data under the endpoint "/monitor-name/...".

The Vue.js component runs on the browser and constructs dynamically the web page of the monitor with the information coming from the monitor-interface. Vue has been chosen for the project because it allows an easy manipulation of the new data loaded by the page and a fast and simple update of the information shown by the interface. HTML and CSS are well handled by the Vue JavaScript libraries limiting the work of the developer and optimizing the result. The component stays in sync with the backend using events passed by a web socket connection.

## V. AUTO-RECOVERY

SEU (Single Event Upset) can happen in the on-detector electronics when radiation or high energy particle passing by the electronic boards flips one of the bits of the memory saved in the registers. Any word composing the event can be reconstructed even if the values that it is bringing are not valuable anymore (they will be discarded in the offline analysis), but it may happen that a SEU breaks the integrity of the event and the DCC board is not able to send it. Usually it brings to a stop the flow of the data and the run goes to error. Radiation can also reset an electronic component of the FE which will lead to a cascade of errors in the off-detector electronics. For example, a reset of a token ring can result in loss of communication between the token ring itself and CCS. On the other end, the FE channels connected to the token ring report a link error to DCC. To avoid these unpredictable interruptions a recovery mechanism has been implemented.

TABLE I
PERFORMANCE OF ECAL AND ES

| Year | LHC Delivered ($fb^{-1}$) | CMS Recorded ($fb^{-1}$) | Lost Luminosity by ECAL wrt LHC (%) | Lost Luminosity by ES wrt LHC (%) |
|---|---|---|---|---|
| 2016 | 41.07 | 37.82 | 0.57 | 0.46 |
| 2017 | 50.25 | 45.39 | 0.30 | 0.09 |

The error caused in the board by a SEU can be detected by the resource supervisor and it causes a state transition of DAQ from "Running" to "Running with SEU". The RCMS LV0 pauses the run (triggers are stopped) and sends a "Recover SEU" message to the FM and consequently to the ECAL Supervisor. Once the message is received the FM changes DAQ to "Recovering SEU" state. The ECAL Supervisor distributes the "Recover SEU" message to all the Resource Supervisors. At this point the software that has detected the problem can try to fix it. The main procedure adopted is based on the reconfiguration of the full electronics or of only part of the registers depending if the error is unknown or well identified. When the recovery action is concluded, the resource supervisor moves to "FixedSEU" and the changed state is transmitted up to LV0 FM. In this way, the triggers are enabled again, and the run can resume.

The main advantage of this system is in transforming a blocking error to a reconfiguration on-the-fly. The run instead of being stopped for ~3-5 minutes to allow the stopping, the full reconfiguration, and the restarting of the system, is paused for ~20-40 s. In this way, the luminosity lost is reduced. The same recovery mechanism used for SEU has been applied to errors arising only in off-detector electronics. For example, ES DCC boards are more sensible to clock perturbation that brings to error in the reading of the FE. In this case when the error is identified, the run is not blocked but the faulty component is reconfigured during the pause of the run.

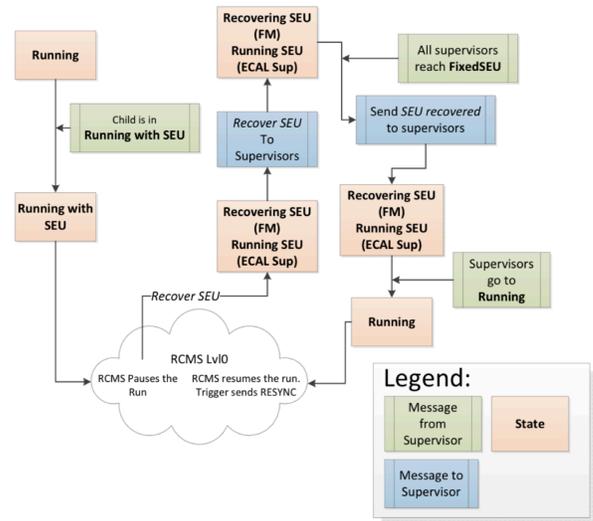

Fig. 6. The schematic for the auto-recovery procedure. The RCMS Lvl0 intervenes when an SEU is encountered and pauses the run. The FM tries to resolve the SEU by communicating with ECAL Supervisor. When the SEU is resolved the RCMS Lvl0 puts the run back in running mode

## VI. CONCLUSION

The auto-recovery process has been found to be successful in reducing the lost luminosity of ECAL and ES and thus increasing their performance. In 2016, the auto-recovery system was able to intercept only a few types of errors. In 2017, the method has been extended in order to react to other kinds of errors like SEU induced resetting of FE electronics. For ES, the improvements in the performances are even more evident since an auto-recovery system like the one presented for ECAL has been deployed only in 2017. Further work on the system has been scheduled for 2018. The upgrade will regard the auto-masking of those channels that cannot be recovered with a simple reconfiguration of the DCC board.

The ECAL monitoring system has been developed to provide a clear overview of the electronics status. The tool is oriented for ECAL DAQ experts. Future work is foreseen to improve the system so that it can also be used by shifters without technical expertise in ECAL DAQ. A summary monitor of the DAQ status will be developed as well as a more advanced access to the error database in order to easily query the history of each electronic board.

ACKNOWLEDGEMENT

I thank Giacomo Cucciati for his guidance and his valuable inputs without which this paper would not have been possible.